# Miniature X-Ray Solar Spectrometer (MinXSS) – A Science-Oriented, University 3U CubeSat

James P. Mason[1] and Thomas N. Woods[2]
*Laboratory for Atmospheric and Space Physics, University of Colorado at Boulder, Boulder, CO, 80303*

Amir Caspi[3]
*Southwest Research Institute, Boulder, CO, 80302*

Phillip C. Chamberlin[4]
*NASA Goddard Space Flight Center, Greenbelt, MD, 20771*

Christopher Moore[5], Andrew Jones[6], Rick Kohnert[7], and Xinlin Li[8]
*Laboratory for Atmospheric and Space Physics, University of Colorado at Boulder, Boulder, CO, 80303*

Scott Palo[9]
*University of Colorado at Boulder, Boulder, CO, 80303*

and
Stanley C. Solomon[10]
*National Center for Atmospheric Research, Boulder, CO, 80301*

**The Miniature X-ray Solar Spectrometer (MinXSS) is a 3-Unit (3U) CubeSat developed at the Laboratory for Atmospheric and Space Physics (LASP) at the University of Colorado, Boulder (CU). Over 40 students contributed to the project with professional mentorship and technical contributions from professors in the Aerospace Engineering Sciences Department at CU and from LASP scientists and engineers. The scientific objective of MinXSS is to study processes in the dynamic Sun, from quiet-Sun to solar flares, and to further understand how these changes in the Sun influence the Earth's atmosphere by providing**

---

[1] Graduate Research Assistant, Aerospace Engineering Sciences, 3665 Discovery Dr., Boulder, CO, 80305, Student Member AIAA.
[2] Associate Director for Technical Divisions, 3665 Discovery Dr., Boulder, CO, 80305, Member AIAA.
[3] Research Scientist, Department of Space Studies, 1050 Walnut St., Suite 300, Boulder, CO 80302.
[4] Research Astrophysicist, Heliophysics Division, 8800 Greenbelt Rd, Greenbelt, MD, 20771.
[5] Graduate Research Assistant, Astrophysical and Planetary Sciences, 429 UCB, University of Colorado Boulder, 80309.
[6] Research Scientist, 3665 Discovery Dr., Boulder, CO, 80305.
[7] Professional Research Assistant, 3665 Discovery Dr., Boulder, CO, 80305.
[8] Professor, Aerospace Engineering Sciences, 429 UCB, University of Colorado Boulder, 80309.
[9] Professor, Aerospace Engineering Sciences, 429 UCB, University of Colorado Boulder, 80309, Associate Fellow AIAA.
[10] Senior Scientist, High Altitude Observatory, 3090 Center Green Drive, Boulder, CO 80301.

**unique spectral measurements of solar soft x-rays (SXRs). The enabling technology providing the advanced solar SXR spectral measurements is the Amptek X123, a commercial-off-the-shelf (COTS) silicon drift detector (SDD). The Amptek X123 has a low mass (~324 g after modification), modest power consumption (~2.50 W), and small volume (6.86 cm x 9.91 cm x 2.54 cm), making it ideal for a CubeSat. This paper provides an overview of the MinXSS mission: the science objectives, project history, subsystems, and lessons learned that can be useful for the small-satellite community.**

## Nomenclature

| | | |
|---|---|---|
| $I_{max}$ | = | maximum current from solar cells, A |
| $I_{Reg}$ | = | current output from regulating buck converter, A |
| $R_{Batt}$ | = | Resistance of battery pack, Ω |
| $R_{CL}$ | = | resistance of current limiting resistor for pseudo-peak power tracking, Ω |
| $R_{S/C}$ | = | spacecraft load, Ω |
| $V_{Batt}$ | = | voltage of battery pack, V |
| $V_{Reg}$ | = | voltage output from regulating buck converter, V |

## I. Introduction

CUBESATS are now becoming a viable vehicle for scientific measurements in space. As commercial entities, government laboratories, and universities continue to miniaturize the requisite technologies for satellites, the sophistication and size of space-based scientific instruments increases. The University of Colorado, Boulder (CU) and the Laboratory for Atmospheric and Space Physics (LASP), developed the Colorado Student Space Weather Experiment (CSSWE, [1, 2]) 3U CubeSat, which launched in 2012 and operated for approximately two years. The science instrument measured high-energy electrons and protons in Low Earth Orbit (LEO) and has resulted in many peer-reviewed journal articles. The present work builds on this success and takes advantage of new commercially available precision three-axis attitude determination and control to achieve fine target pointing toward the Sun. This paper provides an overview of the mission and lessons learned from its development.

## II. Mission Overview

The Miniature X-Ray Solar Spectrometer (MinXSS) is a 3U CubeSat that began development as an aerospace student project at CU and LASP in August 2011. The primary objective of the science-oriented MinXSS CubeSat is to better understand the energy distribution of solar soft X-ray (SXR) emission and its impact on Earth's ionosphere, thermosphere, and mesosphere (ITM). With NSF support in 2013 and subsequent NASA funding in 2014–2016, three MinXSS units have been fabricated (Fig. 1): a prototype and two flight models. The prototype MinXSS has been valuable for early testing and fit checks, and as extra unit for developing flight software in parallel with other build activities. MinXSS flight model 1 (FM-1) is ready for launch and is manifested on an International Space Station (ISS) resupply mission by Orbital Sciences ATK (OA-4), to be launched on an Atlas V on 2015 December 3. MinXSS FM-1 will be deployed from a NanoRacks CubeSat Deployer on the ISS in January 2016, where it will have an expected 5–12 month orbital lifetime, dependent on atmospheric conditions. MinXSS FM-2 is being planned for a higher altitude, longer mission in a sun-synchronous polar orbit (SSPO) via a launch on the Skybox Minotaur C in 2016. This section provides an overview of the science objectives and the history of the project.

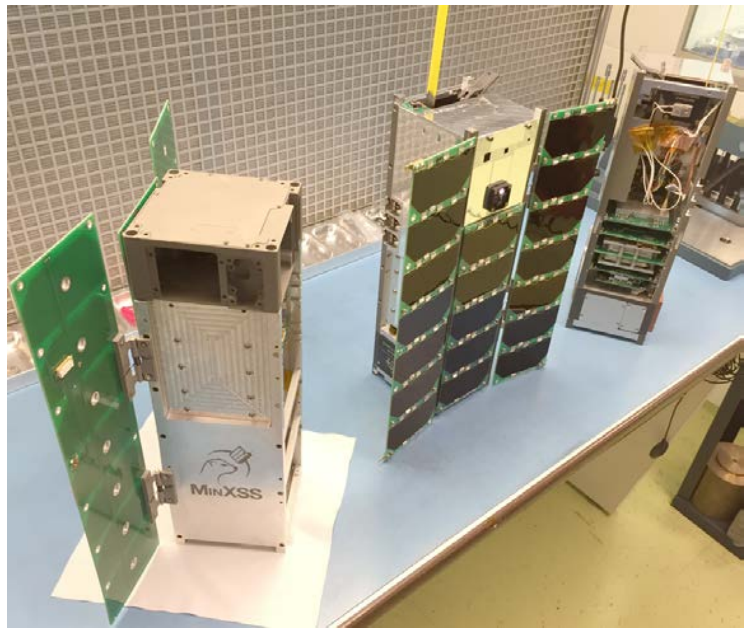

Fig. 1. Photo of MinXSS family (left to right): prototype unit, FM-1, and FM-2.

### A. Science Objectives

There is a rich history of solar SXR spectral observations over the past three decades, but with a significant gap of spectrally resolved measurements in the 0.4–6 nm range (see Fig. 2). There were many new discoveries about solar flares during the 1980s using solar SXR spectral measurements from the DoD P78-1, NASA Solar Maximum

Mission (SMM), and JAXA Hinotori satellites. For example, Doschek [3] provides results about flare temperatures, electron densities, and elemental abundances for some flares during these missions. A review of flare observations from Yohkoh and the Compton Gamma Ray Observatory (CGRO), for the hard (higher energy) X-ray (HXR) range, is provided by [4]. These earlier missions laid a solid foundation for studies of flare physics and flare spectral variability that the Reuven Ramaty High Energy Solar Spectroscopic Imager (RHESSI) [5] and the Solar Dynamics Observatory (SDO) [6] continue today for the HXR and EUV ranges, respectively. With solar flare spectral variability expected to peak near 2 nm [7], in a range not currently observed by any spectrometer, MinXSS measurements of the solar SXR irradiance will provide a more complete understanding of flare variability in conjunction with measurements from RHESSI and SDO EUV Variability Experiment (EVE) [8].

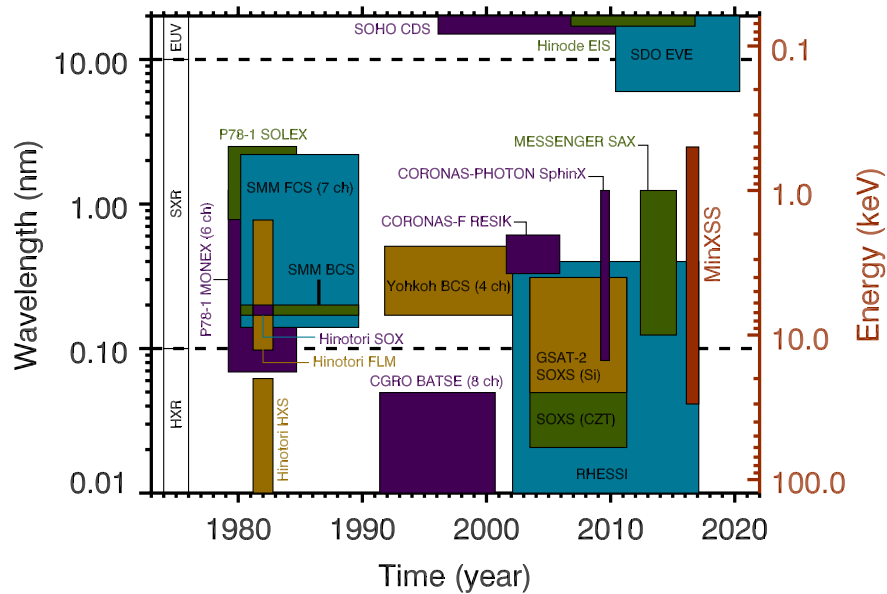

**Fig. 2. History of solar spectral measurements in and near soft x-ray energies (not exhaustive).**

There are also nearly four decades of broadband (5-10 nm wide) SXR measurements not shown in Fig. 2 as they do not provide spectrally resolved measurements. The very limited spectral information from these broadband measurements cannot quantify the specific spectral energy distribution, nor directly quantify the varying contributions of emission lines (bound-bound) amongst the thermal radiative recombination (free-bound) and thermal and non-thermal bremsstrahlung (free-free) continua. These broadband measurements include, among others, the two GOES X-Ray Sensor (XRS) channels covering a combined band of 1.6–25 keV (0.05–0.8 nm) and the even broader band of 0.2–12 keV (0.1–7 nm) from several missions, including the Yohkoh Soft X-ray Telescope

(SXT, 1991–2001; [9]), Student Nitric Oxide Experiment (SNOE, 1998–2002; [10]), Thermosphere-Ionosphere-Mesosphere Energetics and Dynamics (TIMED, 2002–present; [11]), the Solar Radiation and Climate Experiment (SORCE, 2003–present; [12]), and SDO, (2010–present). Broadband measurements of solar SXRs have helped to resolve an outstanding difference between ionospheric models and measurements, such as the electron density from the Haystack Observatory incoherent scatter radar at Millstone Hill. In particular, the SNOE solar measurements were able to resolve the factor-of-4 difference between models and measurements because the SNOE data indicated much more SXR irradiance than had been previously thought [13]. Additional broadband SXR measurements have been made since then; however, differences still remain in understanding solar SXR spectral distribution and atmospheric photoelectron flux. While smaller, these discrepancies are still as large as a factor of 2 at some wavelengths, as shown in Fig. 3; the lack of spectral resolution in the SXR range is thought to be the culprit for most of these disagreements. For example, [14] show that discrepancy between photoelectron measurements and models were significantly improved with new EUV spectral measurements down to 6 nm, and we anticipate further improvement with new solar SXR spectral measurements and atmospheric modeling with data from MinXSS due to its ability to measure all wavelengths in its spectral range simultaneously and with the relatively high spectral resolution of 0.15 keV FWHM.

**B. Solar Flare Studies**

Spectral models of the solar irradiance (e.g., CHIANTI; [15, 16]) are needed in order to convert spectrally-integrated broadband measurements into irradiance units. Detailed modeling to estimate the SXR spectrum during a flare in April 2002 using a set of broadband measurements from the TIMED Solar EUV Experiment (SEE) was performed by [7]. The CHIANTI spectral model is part of their analysis and is also routinely used for processing these broadband measurements (e.g., [17]). While the CHIANTI spectra are scaled to match the broadband SXR irradiance in data processing, there are significant differences for individual emissions lines between the CHIANTI model and observations, often more than a factor of two [18, 19]. Furthermore, there are concerns that CHIANTI could be missing many of the very hot coronal emissions lines, especially in the SXR range where there are so few spectral measurements between 0.5 and 6 nm. Additionally, there are factor of 2 differences when comparing the irradiance results from different broadband instruments, which are worst during times of higher solar activity (Fig. 3). These discrepancies can be partially explained by wavelength-dependent instrument calibrations, but the greater contribution is likely the lack of knowledge of how this dynamical part of the solar spectrum changes as a function of wavelength and time.

The MinXSS spectrometer, an Amptek X123-SDD, flew on the SDO/EVE calibration rocket payload in June 2012, and that measurement had a difference of almost a factor of 8 below 2 nm as compared to the CHIANTI model prediction based on SORCE XPS broadband measurements [20]. This rocket result was a surprise considering that the SORCE-based CHIANTI model prediction agreed with SDO/EVE measurements down to 6 nm. Improvement of models of the solar SXR spectra, which is only possible with calibrated spectral measurements of the SXR emission, is critical to properly interpret these broadband measurements. Our goal with MinXSS observations is to reduce these SXR spectral differences from factors of 2 or more down to less than 30%. In addition, MinXSS will measure solar SXR spectra with higher spectral resolution of 0.15 keV FWHM, as compared to the 0.6 keV FWHM resolution of the most recent analogous instrument, MESSENGER SAX [21]. The MinXSS measurements will enable improvements to solar spectral models, such as CHIANTI and the Flare Irradiance Spectral Model (FISM; [22, 23]). By using MinXSS to improve the FISM predictions in the SXR range, atmospheric studies over the past 30 years will be possible, such as those for the well-studied Halloween 2003 storm period, as well as future space weather events after the MinXSS mission is completed. Getting this spectral distribution of solar flare energy in the SXR range is critical as a driver for atmospheric variations, and will be discussed later in Sec II.D.

The MinXSS data will also help improve understanding of the physics of solar flares themselves. The 0.5–9 keV (0.13–2.4 nm) range observed by MinXSS is rich with high-temperature spectral lines from coronal plasma with temperatures from ~5 to 50 million K, which are greatly enhanced during even small solar flares. MinXSS will also observe the underlying free-free and free-bound continua, extending out to 20–30 keV, which can provide an independent diagnostic of the emitting plasma temperatures. Understanding how solar flares heat plasma, especially up to many tens of million K, is a pressing question in solar physics (e.g., [19, 24, 25]), and the MinXSS observations will provide the best spectral measurements in this energy range to date. Observing the variations of spectral lines in comparison to the continuum will also provide insight into coronal elemental abundances, particularly for Mg, Si, Fe, S, and Ar, to help measure abundances and to understand how they may vary with solar activity and during flares.

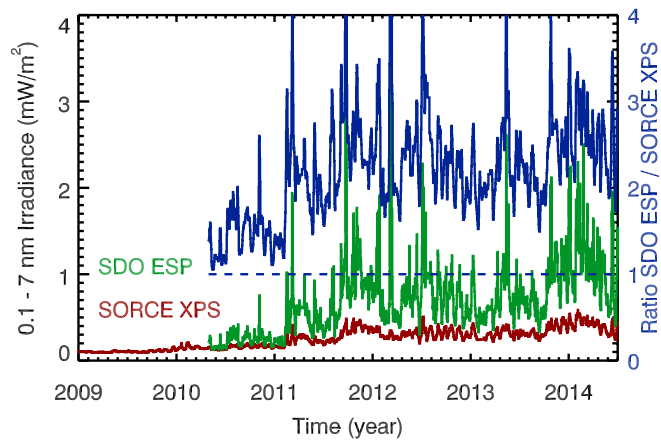

Fig. 3. Solar 0.1–7 nm irradiance currently measured by broadband SXR photometers onboard NASA's SORCE and SDO satellites.

C. Quiescent-Sun Studies

Examples of data analysis and spectral modeling for two quiescent (non-flaring) solar measurements made with the X123 aboard the SDO EVE calibration rocket flights in 2012 and 2013 are provided by [20]. One of the tantalizing results from these two 5-minute observations is that the coronal abundance of certain elements is different for the quieter SXR spectrum on June 23, 2012 than the more active (but not flaring) Sun on October 21, 2013. These abundance differences suggest that different heating mechanisms occur in the quiet network versus active regions, and support the concept that numerous small impulsive events ("nanoflares," e.g., [7, 26]) could be the source of the active region heating. Identifying the mechanism responsible for heating the quiet Sun corona to millions of degrees, while the photosphere below it is only 6000 K, remains one of the fundamental outstanding problems in solar physics [27]. We anticipate that 1–3 months of MinXSS measurements of the solar SXR spectrum will provide adequate data on active region evolution and several flares to more fully address these questions on nanoflare heating. The SXR variability is about a factor of 100–1000 over the solar cycle and can be as much as a factor of 10,000 for the largest X-class flares; MinXSS will be able to observe not only small (A- or B-class flares), but also emission from the truly quiet Sun, as well.

D. Improvements to Earth Atmospheric Models

Energy from SXR radiation is deposited mostly in the ionospheric E-region, from ~80 to ~150 km, but the altitude is strongly dependent on the incident solar SXR spectrum. This wavelength dependence is due to the steep slope and structure of the photoionization cross sections of atmospheric constituents in this wavelength range. The main reason that Earth's atmospheric cross section changes so dramatically in this range is due to the K-edges of O at 0.53 keV (2.3 nm) and of N at 0.4 keV (3.1 nm). Fig. 4 shows two different solar SXR spectra (left) and the result of their absorption in Earth's atmosphere (right). Although the two solar spectra are normalized to have the same 0.1–7 nm integrated irradiance value, their peak energy deposition near the Earth's mesopause has a separation of about 5 km. This separation is considered significant because it is approximately equal to the scale height at 100 km, it is critical to E-region electrodynamics, and the mesopause (the coldest region of the atmosphere) is a critical transition between the middle and upper atmosphere.

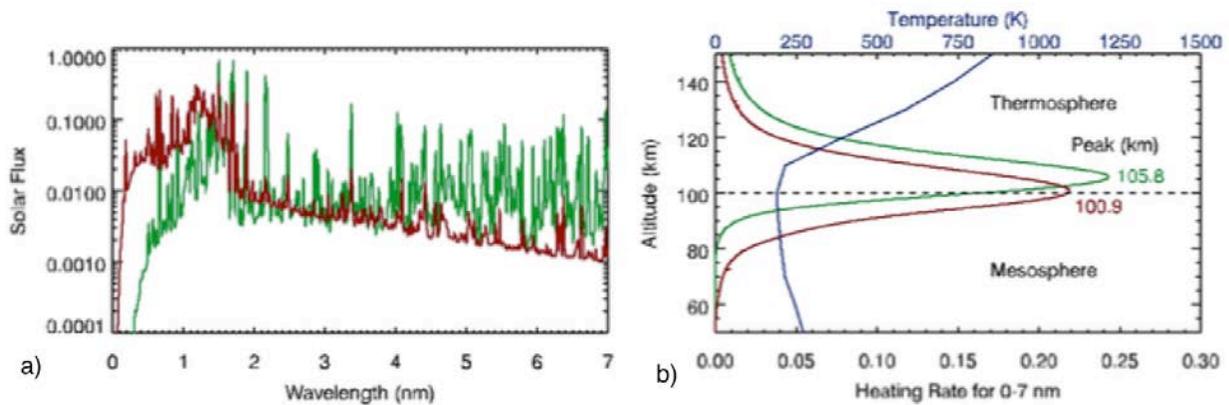

**Fig 4.  a) Two examples of CHIANTI model solar spectra at 0.01 nm resolution, scaled to have identical 0.1–7 nm integrated energy flux: a Sun with bright but non-flaring active regions (*green*), and a solar flare (*red*). b) Earth atmospheric absorption profiles resultant from the two incident solar spectra in a).**

The MinXSS solar SXR spectra are also important to address outstanding issues concerning E-region conductance that has an enormous effect on global electrodynamics and the F-region, especially through the influence of the equatorial electrojet. One of the issues concerns the inability of global general circulation models or detailed process models to produce enough ionization to agree with the E-region peak densities from measurements or well-established empirical models. There appears to be insufficient energy in the solar spectra used as model input, either in the SXR region (especially ~1–3 nm) or at H Lyman-$\beta$ 102.6 nm. The latter has been well quantified by TIMED and rocket measurements. Thus, the focus on the solar SXR spectrum may reveal this missing energy for the E-region. If so, the models could more accurately describe important phenomena such as the magnitude and morphology of the equatorial ionization anomalies, pre-reversal enhancement of the vertical electric field, and the effects of tidal perturbations on the F-region.

**E. Project History and Future Plans**

The MinXSS project began in the Fall semester of 2011 as a graduate student project in the Aerospace Engineering Sciences (AES) department at CU and ran through the Spring 2014 semester, with an average of 11 graduate students each semester who came from various departments: AES, Electrical Engineering (EE), Computer Science (CS), and Astrophysical and Planetary Sciences (APS). The graduate projects course lectures covered general topics such as project life cycle, project management, and systems engineering, and also covered special topics oriented toward spacecraft design such as thermal engineering in vacuum, how to calculate an instrument measurement equation, and an introduction to solar physics. Most students involved in the project have been Masters students, two are including MinXSS-related work in their PhD dissertations, three were undergraduates, and one was a high school student.

The AES department supported the first year of the project. These funds provided the means to obtain mechanical stock and electrical components for the first prototype of our CubeSat Card Cage (see Section 4.A). The National Science Foundation (NSF) awarded limited funds to support the second year of the project, which enabled building of the first full prototype of the CubeSat – including the structure, command and data handling (CDH) custom board, electrical power system (EPS) custom board, custom motherboard, custom battery pack, and plastic 3D printed prototypes of the secondary instrument housing and antenna deployment module. NASA awarded full funding in the project's third year to support the flight build, integration, environmental testing, mission operations, data analysis, and public data distribution. At the present time, flight model 1 (FM-1) has completed environmental tests and is ready for delivery and its launch later in 2015, with deployment expected in early 2016. A second flight unit (FM-2) was built in parallel and is now ready for environmental testing later in 2015 and launch in mid-2016.

Through NASA's Educational Launch of NanoSatallites (ELaNa) initiative, MinXSS FM-1 was manifested on the Orbital Sciences ATK OA-4 launch. The launch vehicle is a United Launch Alliance Atlas V with a Centaur upper-stage and the Orbital Sciences Cygnus to dock with the ISS. The launch is currently scheduled for 2015 December 3. In late January 2016, a NanoRacks CubeSat Deployer attached to the Japanese robotic arm will deploy MinXSS from the ISS at a 45º angle from the ISS – anti-ram and toward nadir – at ~1 m/s. The altitude of the ISS is variable; it periodically boosts to counteract orbital decay due to atmospheric drag. Thus, the precise MinXSS initial altitude will depend on time of deployment, but will be approximately circular at 400 km with an inclination of

51.65º. The precise orbital lifetime for MinXSS depends strongly on solar conditions and the altitude of the ISS at the time of deployment, but we anticipate 5 to 12 months of operations.

MinXSS FM-2 is planned for launch on the Skybox Minotaur C launch in 2016 to SSPO at an altitude of about 500 km. Analysis using NASA's DAS 2.0.2 software was performed to determine the maximum altitude MinXSS can accept while meeting the NASA requirement of "coming down" within 25 years. Assuming a circular orbit and a random tumble (MinXSS active pointing doesn't keep any particular side of the spacecraft in the RAM direction and most of the mission will be a random tumble after the mission ends), any altitude below 620 km will meet the NASA requirement. This second flight of MinXSS will be able to provide longer mission operations, perhaps as much as 5 years, and will incorporate lessons learned from FM-1, e.g., updates to flight software. The primary disadvantage of the SSPO orbit is that it has a harder radiation environment of about 24 kRad over a 5-year mission, as compared to the 2.6 kRad rating for the ISS orbit. The MinXSS processor board has survived radiation tests up to 25 kRad; nonetheless, spot shielding will be added to FM-2 processor and other critical electronics parts prior to delivery.

### III. Mission Architecture

All standard satellite subsystems are present on the MinXSS CubeSat, except for propulsion. Each will be overviewed in the sections below. Fig. 5 shows the requirements flowdown from the science objectives to the mission level requirements, along with the expected performance of the system on orbit. Fig. 6 shows the mechanical block diagram, and Table 1 shows the resource breakdown of the spacecraft subsystems. Volume is only approximate as many components have non-standard geometries. The 4800 g mass limit is derived from the interface control document for the NanoRacks CubeSat Deployer. The more conservative standard mass limit for a 3U CubeSat from the Cal Poly CubeSat Design Specification is 4000 g and would result in a mass margin of 15% for MinXSS.

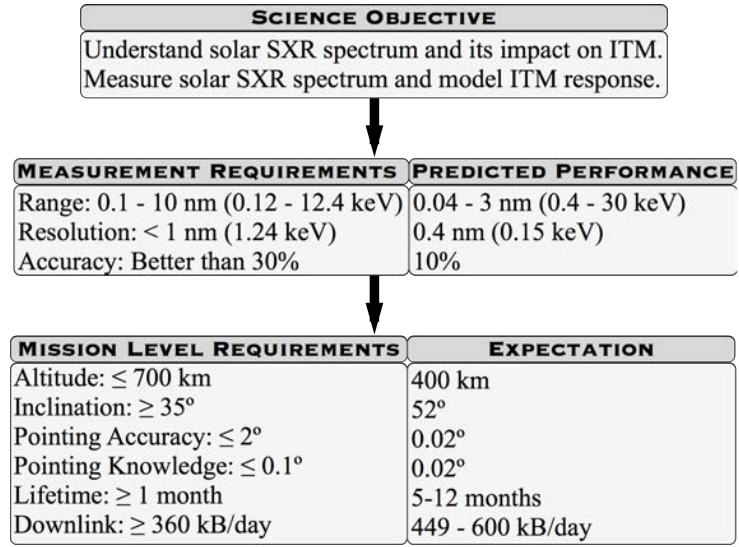

**Fig 5.** High-level requirements flowdown for MinXSS. The measurement requirement for range corresponds to the ISO standard definition for SXRs, and MinXSS is only required to make measurements that fall somewhere within this range. The mission expectations listed are for FM-1 (ISS NanoRacks) only.

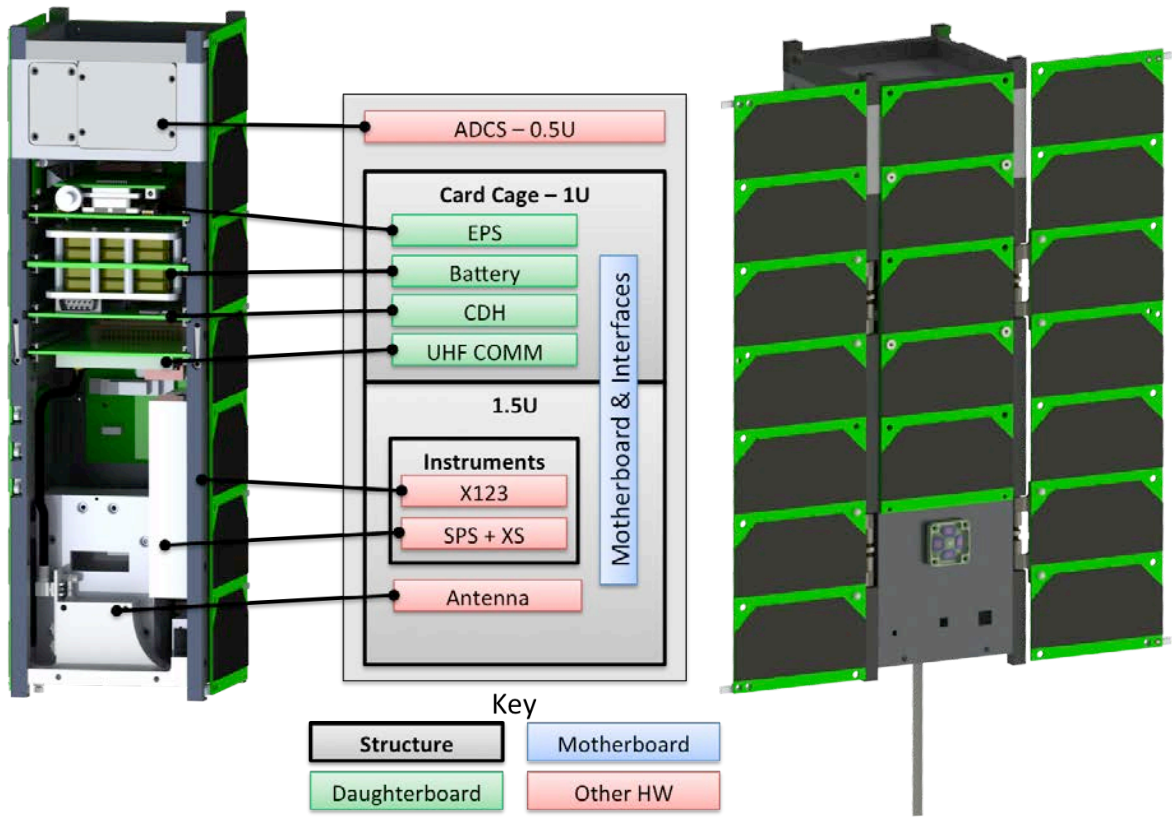

**Fig 6.** MinXSS CubeSat mechanical diagram.

**Table 1. MinXSS CubeSat Resource Breakdown.** Note that under component peak power, not-applicable (N/A) has been inserted into the rows for total, limit, and margin because there is no operational mode where all components will be at maximum power draw.

| Subsystem | Volume [cm$^3$] | Mass [g] | Average Power [W] | Component Peak Power [W] |
|---|---|---|---|---|
| Amptek X123-SDD X-ray Spectrometer | 175 | 323.6 | 2.78 | 5.00 |
| AstroDev Lithium-1 Radio UHF COMM | 144 | 124.6 | 1.49 | 9.80 |
| BCT XACT ADCS | 500 | 870.1 | 1.41 | 1.94 |
| CU CDH | 120 | 46.3 | 0.46 | 0.70 |
| CU EPS + Battery + Solar Panels | 500 | 901.5 | 1.08 | 3.18 |
| CU SPS & XS | 206 | 386.5 | 0.25 | 0.25 |
| CU Thermal Radiator + Heaters | 32 | 2.6 | 0.74 | 3.30 |
| CU Structure + Motherboard | 435 | 856.0 | 0.09 | 0.22 |
| Total | 2264 | 3511.2 | 8.30 | N/A |
| Limit for Volume & Mass, Goal for Power | 3405 | 4800.0 | 10.00 | N/A |
| % Margin | 34% | 38% | 20% | N/A |

**A. Primary Instrument – Amptek X123-SDD**

The purpose of the primary MinXSS science instrument is to measure solar spectra within the ISO standard SXR range of 0.1–10 nm range listed in the requirements flowdown (Fig. 5). In order to function within a CubeSat, the instrument must be low mass, low power, and have a small volume. A commercial-off-the-shelf (COTS) solution perfectly met these design requirements. The Amptek X123-SDD weighs ~324 g after custom modifications were made for mounting to the CubeSat and thermal foam was added for cooling electrical components in vacuum. It consumes approximately 2.5 W of power nominally, and 5.0 W for approximately 1 minute when first powered on. Much of the power draw (including the initial transient) results from the integrated thermo-electric cooler (TEC) reducing the temperature of the SDD to the user-defined set point (–50º C for MinXSS). The dimensions of the X123 are also sufficiently small to easily fit within a CubeSat due to the manufacturer's designed purpose as a handheld SXR measurement unit for geological fieldwork. The X123-SDD's ~500 µm active thickness and ~16 µm beryllium (Be) entrance window define a spectral range sensitivity of ~0.4 keV to 30 keV (~0.04–3 nm), which covers the primary range of interest for scientific studies of 0.5–2 nm. The instrument includes all the necessary processing electronics, including an integrated multi-channel analyzer, to produce a spectrum that is output via an RS232 interface. It can also be commanded programmatically to change numerous parameters such as integration time and energy thresholds. The custom modifications for spaceflight include staking the larger electronics components, adding a mounting plate for the electronics, adding a custom interface cable and 9-pin connector, adding a tungsten plate with pinhole aperture for the SDD, and providing stainless steel radiation shielding around the aluminum detector vacuum housing.

In October 2014, the MinXSS science instruments, including the X123, were calibrated at the National Institute of Standards and Technology (NIST) Synchrotron Ultraviolet Radiation Facility (SURF; [28]). The synchrotron radiation provides a calibrated continuum emission source, with a radiometric accuracy of ≲10% in the SXR range. The SURF electron storage ring beam energy is adjustable from 60 MeV to 416 MeV; the synchrotron spectral distribution is dependent on the beam energy, and the MinXSS calibrations use the higher beam energies in order to maximize the incident SXR flux. The absolute radiometric calibration of the X123, as a function of wavelength, is then obtained by comparing the measured output spectra with the known incident photon flux from the SURF beam; an example, and further description, can be found in [20]. The narrow spatial extent of the SURF beam in the X-ray range allows for a mechanical determination of the instrument optical axis ("boresight") relative to a reference frame, and the uniformity of response over the instrument's ±4° field of view (FOV) is determined using a gimbal system to rotate the detector optical axis about the incident beam. The nonlinearity of the detector electronics is measured by adjusting the intensity of the incident synchrotron flux.

**B. Secondary Instrument – SPS & XS**

The purpose of the secondary instrument is to provide support for scientific analysis of data from the primary instrument. Two sensors are needed to achieve this: one to provide independent high-precision attitude knowledge of the solar position and another to provide an in-flight SXR irradiance reference. Again, these instruments must be low mass, low power, and small volume to be accommodated within a CubeSat platform. MinXSS heavily leveraged instrument heritage from the larger GOES-R EUV X-ray Irradiance Sensor (EXIS) development at LASP, which already met all of these requirements. The custom-designed Application Specific Integrated Circuit (ASIC), in particular, provides the backbone of this exceptionally low power, low noise system. A custom mechanical design for the casing was necessary to integrate the subsystem with MinXSS, which was manufactured for flight using aluminum sintering (3D) printing.

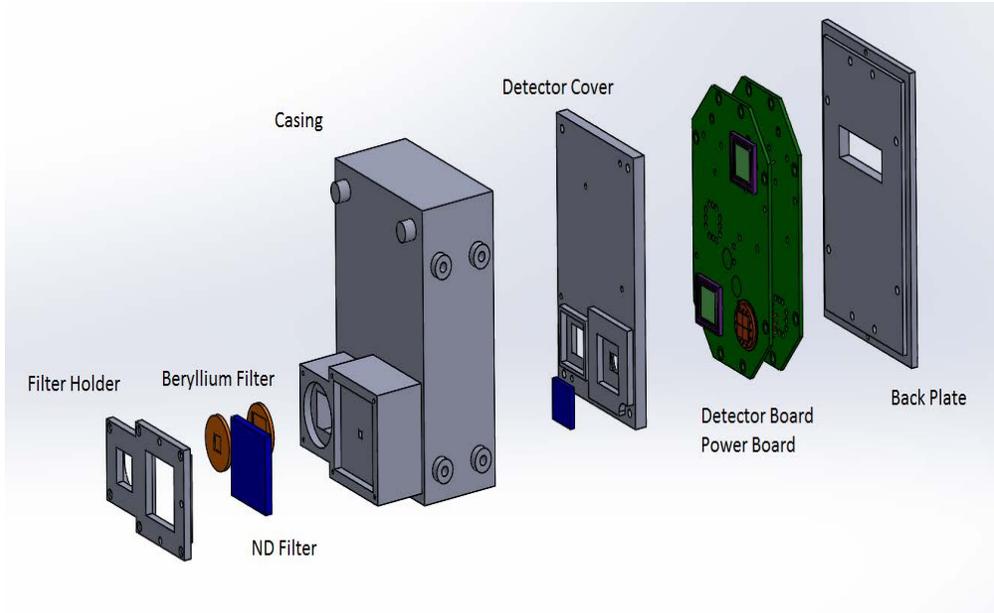

Fig. 7. Solar Position Sensor and X-ray Sensor (SPS & XS) exploded view.

Fig. 7 shows an exploded mechanical view of this secondary instrument. The Solar Position Sensor (SPS) is a quad-diode with effective neutral-density-7 filter and 2 mm × 2 mm square knife-edge aperture, with an FOV of ±4º. The solar visible light falls on the four diodes such that the illumination on each diode depends on the incoming angle of the solar radiation. The resultant measurements are used to compute the Sun's position to better than 1 arc-min (3σ) as described by [29]. These data are sent to the attitude determination and control system (ADCS) for inclusion in the fine-attitude control solution, and telemetered to the ground for use in science processing. The X-ray Sensor (XS) is a single diode with two Be foil filters, whose total ~16 µm thickness is matched to the X123 to define a response over the same ~0.04–3 nm wavelength range. XS has a 5.0 mm diameter knife-edge aperture and an FOV of ±4º. The diode operates in photocurrent mode, integrating the total SXR flux over its bandpass and integration period; this provides a measurement that can be compared to the integrated X123 spectrum, to within measurement and calibration uncertainties. These data are also telemetered to the ground for use in science data processing.

The SPS & XS were also calibrated at NIST/SURF. The SPS optical axis and the transfer equation relating off-axis position to quad-diode output were determined using the gimbal system to rotate the optical axis around the incident SURF beam. The XS optical axis and uniformity of response over its FOV were similarly determined. The absolute radiometric response of the XS was determined similarly to the X123, comparing the known incident synchrotron photon flux with the output from the photodiode. (No absolute calibration was necessary for the SPS.)

The SPS & XS system, including ASIC, had been previously measured to be highly linear through testing during GOES-R development, so the MinXSS calibrations omitted nonlinearity testing.

### C. CDH and Flight Software

The core of the MinXSS Command and Data Handling (CDH) subsystem is a low-power Microchip dsPIC33 Microcontroller Unit (MCU, MC dsPIC33EP512MU810). The CDH communicates with and controls the X123 instrument, UHF COMM, and ADCS via RS232, monitors voltages, currents, and temperatures via $I^2C$ for the motherboard, CDH, COMM, EPS, and SPS & XS, and reads detector data from the SPS & XS ASIC via digital input/output (DIO). Additionally, the CDH handles all incoming commands, housekeeping monitoring, data manipulation for downlinking data packets, power switching of subsystems, and configuration of the operation modes. Most of the CDH operation is configurable via uplinked command, and several of these CDH processes are autonomous for maintaining a safe power configuration. Data are stored on a 4 GB Secure Digital (SD) memory card, and each type of data packet has its own dedicated circular buffer on the SD card. This SD card can store more than 1400 days (3.8 years) of science, housekeeping, and log message data packets, and 48 hours of ADCS high-rate data packets. The dsPIC33 internal real-time clock (RTC) and an external RTC IC provide precise time knowledge. The external RTC IC also has an EEPROM for storing start-up configuration parameters, which can be modified via uplinked commands. The dsPIC33 watchdog timer is used to initiate a reset of the system in case it becomes unresponsive, and a reset command can also be sent from the ground. The MinXSS FM-1 1-year mission worst-case radiation dose estimate is 2.6 kRad, with a minimum shielding of 2 mm of Al provided by the CubeSat structure. Two of the prototype CDH boards successfully passed radiation tests of 10 kRad and 25 kRad.

The embedded flight software is built on a Slot Real-Time Operating System (RTOS), written in C, as originally developed at LASP for the SDO/EVE rocket experiment. The key elements of the software design are robustness and simplicity, with the health and safety of the satellite as top priority. Because many of the tasks performed by the CDH are not time-sensitive and can be handled at any time in the slot process, the real-time demands on the CDH and flight software are very low. The RTOS uses the dsPIC33 timer with 1 msec resolution for execution of tasks, but most monitoring by the CDH has a cadence of 1 sec or slower.

### D. EPS, Battery, and Solar Panels

The MinXSS electrical power system (EPS) is largely based on heritage from the successful CSSWE direct energy transfer (DET) design. The EPS uses high-efficiency buck converters for power regulation to 3.3 V and 5.0

V and a simple battery charging logic for use with Li-polymer batteries. Minor design modifications were incorporated to accommodate the higher power generation and consumption on MinXSS as compared to CSSWE, as well as more voltage and current monitors. Two additional major differences were implemented: pseudo-peak power tracking (see Section IV.D) and additional switches to prevent the system being powered prior to deployment in order to comply with NanoRacks ISS human-safety standards.

The battery pack consists of four SparkFun 2-Ah Li-polymer batteries, configured as two parallel sets of two batteries in series ("2s2p") to provide a 6-8.4 V unregulated 4-Ah bus, two temperature sensors and two heaters, that are sandwiched between the batteries. Heat transfer tape was used between each layer of the battery pack to achieve a homogenous temperature distribution during flight. The PCB in the middle of the pack does not have a copper plane in its center as all other daughterboards do – the intent being to thermally isolate the batteries from the rest of the system. This was a part of the passive thermal design to create a battery-dedicated thermal zone, as the batteries have the narrowest operating temperature range of all components in the system. Finally, the pack was encapsulated with aluminum plates on standoffs, providing sufficient volume for the batteries to expand under vacuum and thermal cycling. Arathane 5753 with Cabosil glass beads was placed between the batteries and these encapsulation elements to act as a soft bumper to expanding batteries.

MinXSS uses 19 triple-junction GaAs, 30% efficient solar cells from Azur Space Solar Power GmbH. One five-cell solar panel is fixed to the body of the CubeSat on the solar-oriented side, and two seven-cell solar panels will deploy by command to have the same solar-orientation as the body-fixed panel. Because MinXSS is a Sun-pointed spacecraft, these solar panels can nominally supply 22 W at end-of-life during the orbit insolation period. A 100-hour mission simulation test with the fully integrated spacecraft connected to a solar array simulator under various eclipse periods was performed to verify that there is adequate margin for operating all MinXSS subsystems and for charging the battery (see Sec. IV.E). Additionally, flight software incorporated the ability to autonomously power off the X123 (the largest power consuming subsystem) and the other noncritical subsystems during eclipse if there are any battery power issues for eclipse operations. The power-cycling flags can be enabled via command, but we do not anticipate the need for their use.

**E. Communications**

MinXSS leveraged heritage from the CSSWE CubeSat by using the same radio and ground station for UHF communications. The ground station is located on the roof of the LASP Space Technology Research building in

Boulder. It consists of a pair of M2 436CP42 cross Yagi antennas, each with a gain of ~17 dBdc and a circular beamwidth of 21º. A Yaesu G5500 azimuth-elevation rotator controlled by SatPC32 points the antenna system. SatPC32 also accounts for Doppler shifts via its control of the ground radio, a Kenwood TS-200. The antennas and motors are mounted on an ~2.4 m tower and are connected to the electronics in the control room below by ~60 m low-loss cabling, which accrues -5.4 dBm of RF signal loss. The flight radio is an Astronautical Development LLC Lithium-1 radio that operates in the UHF band at 437 MHz. Additionally, the antenna is nearly identical to that used on CSSWE, which is a deployable spring steel tape measure with a length of 47.6 cm. The gain pattern was measured using the MinXSS prototype in an anechoic chamber at First RF Corporation in Boulder, CO. The measurements were compared to a FEKO model and propagated through Satellite Tool Kit (STK) to estimate the expected daily average downlink data capacity: 600 kB/day using the FEKO model or 449 kB/day using the measurements. These estimates are not highly precise due to the limited fidelity of the model and the prototype structure, but provide an idea of what to expect. The requirement of at least 360 kB/day appears to be easily satisfied.

**F. ADCS**

In order to provide a stable view of the Sun for the science observations and to maintain appropriate antenna orientation during ground contacts, MinXSS has an active Attitude Determination and Control System (ADCS). With the wide field of view of the X123 (±4°), the pointing requirements for MinXSS are only 2° (3$\sigma$) accuracy and 0.1° (3$\sigma$) knowledge.

The commercial CubeSat ADCS onboard MinXSS is a fleXible ADCS Cubesat Technology (XACT) from Blue Canyon Technologies (BCT). BCT has developed a 0.5 U-sized ADCS unit (0.85 kg) utilizing miniature reaction wheels, torque rods, a star tracker, a coarse sun sensor, inertial measurement units, and magnetometers. The BCT XACT is expected to provide pointing accuracy and knowledge of better than 0.003° (1$\sigma$) in 2 axes, corresponding to the plane of sky of the star tracker, and 0.007° (1$\sigma$) in the 3rd axis, parallel to the star tracker optical axis. The XACT interface utilizes 5 V and 12 V power inputs (1.0 W nominal, 2.8 W peak) and serial communication (RS232 for MinXSS, but other options are available). SPS provides 2-axis (pitch/yaw) pointing knowledge on the Sun to better than 1 arc-min (3$\sigma$), which can be sent to the XACT for closed-loop fine-Sun pointing; however, the XACT system can easily meet the MinXSS pointing requirements without this additional knowledge.

After integration with MinXSS, multiple tests were performed to verify functionality and performance of the ADCS. A custom air-bearing table was built to provide a relatively torque-free environment for the ADCS to control the spacecraft. For example, we verified that the spacecraft can track the Sun with a heliostat at LASP, that magnetometers reversed sign when the spacecraft was rotated 180º in each axis, that torque rods produced a measurable magnetic field, and that the star tracker took interpretable images and found matches to stars in its library when observing the night sky.

**G. Thermal Design**

In normal operations, MinXSS has the +X side facing the Sun and the –Y face pointing toward deep space (see Figure 8 for axes definition). Thermal Desktop analysis shows that this configuration easily satisfies all component operational and survival temperature requirements. Fig. 8 shows the Thermal Desktop model result for MinXSS FM-1 in the longest eclipse orbital case. Thermally-isolating washers (0.094 cm thick Delrin) are used for mounting the body-fixed solar panels so that, despite solar panel temperatures swinging between –20 ºC and +75 ºC, the components in the system remain within their temperature requirements. All sides of the spacecraft not facing the Sun are radiators. The outer bare aluminum faces of the structure were covered with silver-coated Teflon tape – a high emissivity material. These radiator plates remain cold at most orbital positions, ranging from –16 ºC to 17 ºC. The model was validated using a dedicated thermal balance test after the completion of thermal vacuum environmental testing. The details of the thermal balance test and comparison of those results with the thermal model are the subject of a forthcoming paper.

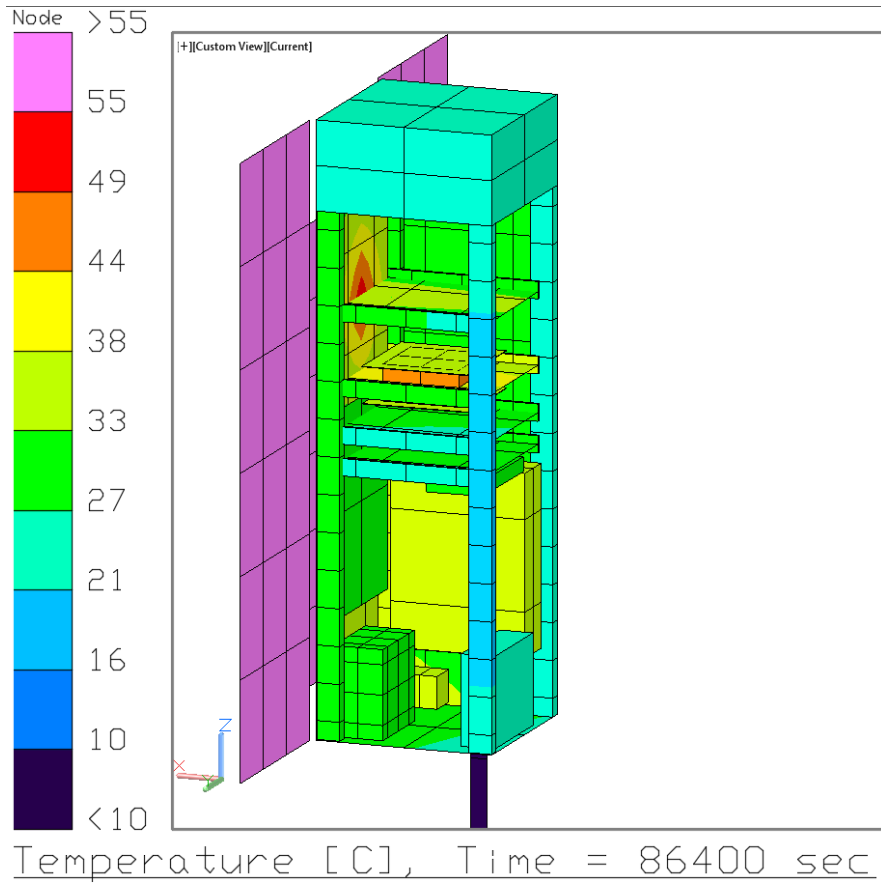

**Fig. 8. Thermal Desktop model result for a single point in time during ISS orbit at β = 0º, with the –X and +Y plates hidden to allow sight of the internal components.**

Temperatures are actively controlled in three places in the system: the battery pack, X123 detector, and the inside wall of the –X structural plate. Following the successful implementation of CSSWE's battery heaters, the two battery heaters on MinXSS trigger when the temperature falls below +5 ºC and deactivate at +10 ºC. The battery heater power, as predicted by Thermal Desktop, is 0.5–0.8 W, orbit-averaged, depending on the orbit β angle. Secondly, the TEC in the X123 detector head can maintain a temperature differential of up to 85 ºC, which was verified under vacuum at LASP. The requirement on measurement noise translates to maintaining the X123 detector head at –50 ± 20 ºC. Thus, the warm side of the TEC must be kept below +40 ºC. This is easily satisfied by strapping the TEC's warm side to the –Y radiator plate, which is always below +17 ºC. Finally, the structure heater was found not to be very effective at heating the subsystems (e.g., CDH, SPS & XS), so it probably will not be used during flight. Nevertheless, the thermal analysis and testing suggests that the combination of passive and active temperature control in MinXSS will be sufficient to maintain all components within their operational temperature limits.

The battery and structure heaters are controlled by CDH. The autonomous control of heaters can be enabled or disabled and the temperature set points can be changed by command. The X123 TEC is controlled by the X123 electronics and its set point can be adjusted by command, but it is always enabled if X123 is turned on.

## IV. Advancing CubeSat Technologies and Lessons Learned

### A. CubeSat Card Cage

Experience with the PC104 PCB interface on the CSSWE CubeSat led the MinXSS team away from the card stack design due to the difficulty in debugging boards once integrated. Instead, the CubeSat Card Cage design uses a motherboard/daughterboard architecture that allows any individual card to be easily removed, and an extender board optionally inserted to have access to the daughterboard for probing while still electrically connected (Fig. 9). Additionally, the standard electrical interface allows boards to be swapped to any position. MinXSS utilizes a DIN 48-pin connector for the daughterboard-motherboard interface. This relatively large connector was chosen for ease of soldering for new engineering students and because it easily satisfied the requirements on the number of necessary pins and mechanical dimensions. In the future, a higher density connector with potentially more pins could be chosen to provide a lower mass and lower volume solution while still providing the flexibility of the card cage architecture.

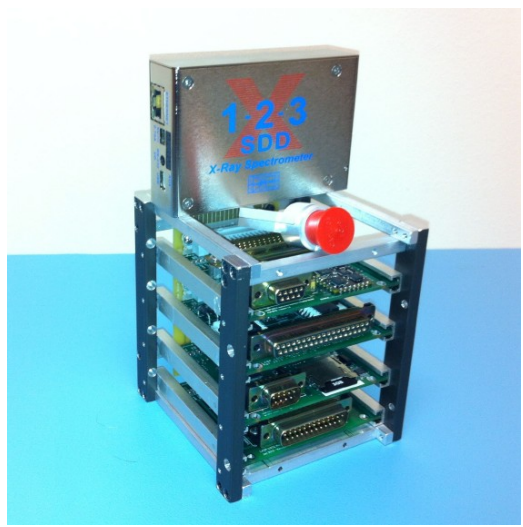

**Fig. 9. Prototype CubeSat Card Cage design.**

### B. 3D Printed Parts

The MinXSS project used 3D printed parts for both prototyping and flight components. For prototyping, the SPS & XS housing was 3D printed in plastic twice as the design iterated, and the solar array hinges were printed in plastic once. This was done using CU's Objet 30 printer with VeroWhitePlus plastic. For flight, these same components were 3D printed in metal using direct metal laser sintering at GPI Prototype. The SPS & XS housing is aluminum with a shot blasted finish (Fig. 10). This finish was very rough and required significant sanding to get an acceptable surface finish and clean edges. The solar array hinges are stainless steel with a shot blasted finish (Fig. 11). A minimal amount of sanding was required for these parts because the requirements were looser and the finish was slightly better than SPS & XS. The better finish was likely due to the hinges being a simpler part that required no filler material during the 3D print (sintering) process.

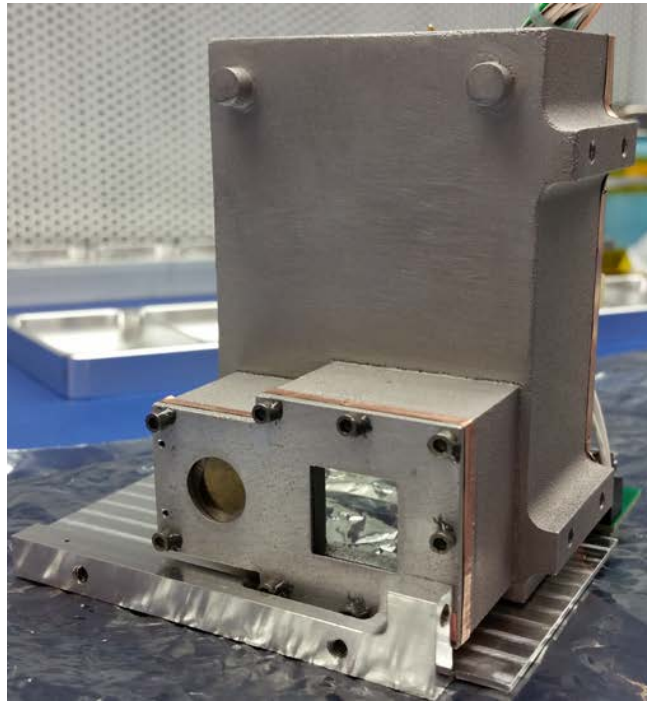

**Fig. 10. Aluminum 3D-printed SPS & XS housing after sanding and integration.**

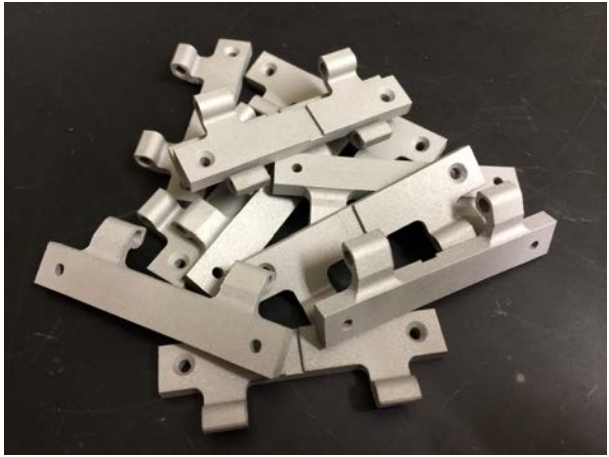

Fig. 11. Stainless steel 3D-printed solar array hinges as delivered from vendor.

As plastic 3D printers become more pervasive, affordable, and precise, the draw toward using the resultant parts for flight is becoming stronger. A major risk that must be addressed is the unknown properties of these materials, particularly in their response to vacuum and UV exposure. We would like to see an open database where specifications based on test results for common 3D print materials, such as ABS and PLA, could be accessed.

C. Simplification of Solar Panel Fabrication Process

CSSWE used epoxy (Arathane 5753) on the back of solar cells to adhere them to the solar panel PCBs. This technique is typical but requires significant assembly and curing time. MinXSS used double-sided Kapton tape with acrylic adhesive to adhere solar cells to the PCBs. We used a specialized rubber vacuum sealer to apply pressure to the cells uniformly and meet the manufacturer's recommended application pressure. This reduced the time to produce a solar panel from three days to one day. To get electrical conductivity from the back of the solar cell to the PCB, we applied silver epoxy in large vias behind each cell. We also tested a new-to-market tape: 3M Z-Axis tape. This tape is electrically conductive between the adhesive and back side and could save the extra step of applying the silver epoxy or soldering/welding on tabs. For flight, Kapton tape was used because 1) the Z-Axis tape adhesive was not rated for as wide a temperature range as the Kapton acrylic adhesive, 2) there was concern that the Z-Axis tape could not sustain the high current of the solar cells for as long as solder or silver epoxy could, and 3) the Z-Axis tape thermal conductivity properties were not specified in the datasheet.

In the future, we would like to see solar cell manufacturers adopt a standard form factor compatible with CubeSats. MinXSS uses 40 mm × 80 mm cells from Azur Space (Fig. 12), which are a great fit within the rail boundaries of CubeSats (maximum of 83 mm wide and 340.5 mm long for 3U CubeSat). The 80 mm width for cells provides 1.5 mm margin on each side from the rails. If the spacing between cells could be reduced to 4.5 mm or less, then there could be 8 Azur Space solar cells instead of 7 on a 3U panel. Alternatively, if the height of the cells were changed to be 50 mm instead of 40 mm, then they would be more modular for fitting one solar cell per 0.5U of the panel length. With six 50 mm x 80 mm cells instead of seven 40 mm x 80 mm cells, there could be 7% more power per 3U panel.

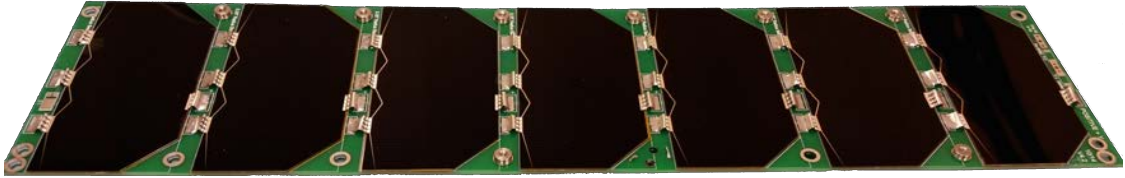

**Fig. 12. A populated 7-cell deployable solar array for MinXSS FM-1.**

**D. Pseudo-Peak Power Tracking**

A modified DET EPS design was implemented on MinXSS that was inherited from the CSSWE CubeSat to include an additional specially selected resistor to create a pseudo-peak power tracking (PPPT) system. The extra resistor was chosen to prevent a rapid voltage drop from the solar cells when the battery attempts to draw a large current, namely when the battery state of charge is relatively low right as the spacecraft exits the orbit eclipse.

In the CSSWE and MinXSS EPS design, the output of the solar panels power 8.6 V regulators that then provide regulated 8.5 V power directly to the battery and system. In this DET design, the batteries will charge up to 8.5 V, and there are no supporting electronics required to control the battery charging process. In reality, this simple approach only provides about 50% of the power intended from the solar panels when the battery capacity is low. In particular, when the battery needs more power input (high current) for charging, the high current draw from the solar cells results in much lower voltage, following the standard solar cell current-voltage (I-V) curve. When the solar panel output voltage goes below the minimum input voltage level of the 8.6 V regulator, the regulator turns off. Consequently, the current drops and the solar panel output voltage increases, and the 8.6 V regulator turns back on. This results in a high-frequency on-off regulator oscillation that had the EPS 8.6 V regulators on for only about 50% of the time during the early part of the orbit dayside during mission simulations. The MinXSS solar panels were designed for 80% of peak efficiency at EOL, but the 50% decrease in power was an unacceptable power loss for the nominal power budget.

The solution for MinXSS, without having to redesign or rebuild the EPS board, was to replace the sense resistor on the output of the solar panel regulator with a larger resistance so that the effective current draw out of the solar panel would be limited and thus would not cause the regulator to turn off. We refer to this current-limiting resistor for the solar panels as pseudo-peak power tracking (PPPT). Fig. 13 shows a simplified version of the PPPT circuit for the MinXSS EPS.

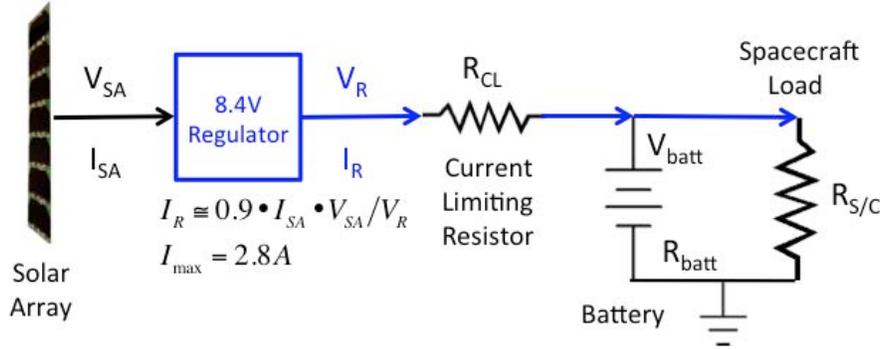

**Fig. 13. A simplified circuit diagram of PPPT used for the MinXSS EPS.**

The value for this current-limiting resistor was estimated for the MinXSS power configuration using Eq. 1. The first term on the right-hand side of Eq. 1 is the current for the spacecraft load, and the second term is the current for charging the battery. The spacecraft load is assumed constant, but the battery charging current starts off high when the battery voltage is low and then ramps down to zero when the battery voltage is the same as the regulator voltage downstream of the current limiting resistor. The ideal value for the current-limiting resistor, $R_{CL}$, is such that it limits the current out of the regulator, $I_{Reg}$, to be less than the maximum current, $I_{max}$, possible from the regulator (at the peak power part of the solar panel I-V curve) and when the battery voltage, $V_{batt}$, is at the lowest allowed level. For the MinXSS design and configuration, the regulator voltage, $V_{Reg}$, is 8.5 V, the worst-case system load (largest power) has 7.0 Ω for $R_{S/C}$, a battery impedance of 0.125 Ω, and a value of 2.8 A for $I_{max}$. The goal for MinXSS was to keep the battery voltage above 7.1 V at all times, so an $R_{CL}$ of 0.25 Ω is the desired value for the MinXSS configuration to satisfy Eq. 1. That is, with this value of $R_{CL}$, $I_{Reg}$ equals $I_{max}$ when $V_{batt}$ equals 7.1 V.

$$I_{Reg} = \frac{V_{Reg} - I_{max}R_{CL}}{R_{S/C}} + \frac{V_{Reg} - I_{max}R_{CL} - V_{Batt}}{R_{CL} + R_{Batt}} \qquad (1)$$

After the current-limiting resistor was installed into the EPS, additional mission simulations were run. We verified that the prediction of the regulator current, $I_{Reg}$, and the measured battery voltage agreed with the measured regulator current.

One disadvantage to the PPPT implementation is that there is additional heating of the EPS board because of the larger resistance; however, this extra heating peaks right after exiting eclipse, the precise time when temperatures are cooler and heating is desired anyway. For example, the power loss (heating) in the PPPT current-limiting resistor is estimated to be 2.6 W when the battery voltage is at its lowest value of 7.1 V, decreasing to 0.93 W when the battery voltage is at 7.5 V, and reduces to less than 0.1 W once the battery voltage is above 8.0 V. The primary caveat in the PPPT design is that resistor tuning must be done a priori, and is fixed, whereas maximum PPT (MPPT) systems can tune resistance in real time to maintain the maximum power point on the solar cell I-V curve. The trade studies performed for CSSWE and MinXSS resulted in the selection of a custom DET EPS due to the simplicity of design. Both teams were unaware of the consequential loss of power generation at the time of the original designs. The advantage of the PPPT circuit is that it is only minimally more complex than DET, adding little risk for a large benefit.

In the future, we would like to see a standard MPPT IC for interfacing to common CubeSat battery packs (e.g., 8.4 V Li-polymer battery packs). We found it difficult to identify a commercial MPPT IC or proven MPPT circuit that could be integrated with our system. We purchased the most promising MPPT IC, a Linear Technology LT3652, and spent significant time attempting to integrate it with the MinXSS EPS, but its intended use prevented proper functioning for our solar panel and battery configuration.

**E. Importance of Flight-like Testing**

Various tests were performed on MinXSS that were geared toward simulating the orbital environment and flight-like operations. These included low-external-torque tests of the ADCS, thermal vacuum with a long-duration mission simulation, early-orbit end-to-end communication testing performed several miles away from the ground station, and detailed battery characterization of the actual batteries to be flown.

Using a custom-built air-bearing table, we tested the functionality and performance of the ADCS. This test simulated an orbital environment with reduced external torques present. Through this testing, we discovered that an operational amplifier (op-amp) was preventing the XACT coarse sun sensor from being properly read by its internal flight software, and this op-amp was replaced to resolve this issue. It is unlikely this would have been discovered otherwise, and may have resulted in the spacecraft not being able to quickly find or accurately track the Sun on orbit. Significant effort in mission operations may have been able to salvage the mission in that situation, but only minor effort was required to replace the offending op-amp. Air-bearing testing requires very careful balancing of the system and as much reduction of external torques as possible; e.g., even air flow from building ventilation could limit the tracking duration while operating on the air-bearing table. It also requires the computation of moments-of-inertia specific to the air-bearing-CubeSat system to be provided to the ADCS for appropriate control to be implemented. Without such an update to the ADCS software, the ADCS response is too sluggish (slow) to confirm that the ADCS is tracking as expected.

Thermal vacuum tests are irreplaceable for determining if the CubeSat can function in vacuum and for measuring performance near the operational limits of components. Through such testing of MinXSS, we discovered a short in a battery heater that reset the entire system every few seconds, which only manifested under vacuum. This was caused by the battery expansion, which created an unintended electrical connection between the two nodes of the heater. Typically, CubeSats are only required to bake out, not perform a functional thermal vacuum test, but we highly recommend this test as a process to increase the success rate of CubeSats.

A 100-hour mission simulation test was performed on MinXSS during four of the eight hot-cold cycles of the thermal vacuum testing. A solar array simulator, with an I-V curve programmed to model the Azur Space solar cells used on MinXSS, was jumpered into the MinXSS EPS board. The jumper bypassed the two deployable solar panels. The output of the solar array simulator was programmatically cycled in intervals corresponding to ISS orbit insolation/eclipse periods at three different β angles. The total orbit period was 93 minutes and the three eclipse periods were 28 minutes (average β), 38 minutes (β = 0º), and 0 minutes (β > 76º). Power performance data was collected for the entire system throughout each of these scenarios, and verified that the PPPT maintained a power positive state through many orbits. Additionally, this test was used to verify the functionality of a flight-software commandable flag to disable power to the X123 during eclipse periods. This option was introduced into the flight software early in the project in anticipation of a marginal power balance. The X123 was chosen for power cycling because it is the largest consumer of power and because the primary science target – the Sun – is not visible in eclipse. However, this is not the default state in the mission design as it introduces excessive power cycling on the primary science instrument; nominal operations leave the X123 powered on during the entire orbit. As the spacecraft performance degrades on orbit (e.g., solar cell efficiency loss), it may become necessary to enable the X123-eclipse-power-cycling flag. Finally, the 100-hour mission simulation test included periodic stored-data downlinking with durations equivalent to the ground station contacts expected on orbit. The 100-hour mission simulation test was the most flight-like testing possible with the facilities available, and greatly increased confidence in and understanding of the system as it will behave on orbit. It also ensured that the flight electronics are likely past the "infant mortality" phase.

End-to-end testing was also performed on MinXSS to verify functionality of the full communication pipeline. The spacecraft was taken several miles away to a position in the line-of-sight of the ground station, and early-orbit commissioning tests performed. This boosted confidence in several areas: that we would meet the NanoRacks requirement of not deploying the MinXSS antenna or solar arrays in the first 30 minutes after deployment from the ISS, that those deployments would be successful, that communications could be established after antenna deployment, and that our ground software commissioning scripts could autonomously perform telemetry verification and commanding.

Significant battery testing was performed to comply with requirements flowed down from NASA Johnson Space Center through NanoRacks to all CubeSats going to the ISS. These requirements are in place to protect astronauts on the ISS and far exceed the standard CubeSat requirements in the Cal Poly CubeSat Design Specification. Nevertheless, we recommend that all CubeSats perform several of these tests, if only to better understand the actual batteries to be flown, i.e., not just batteries from the same lot or of the same type. We found the following to be the most useful tests: visual inspection for dents or leaks, measuring the open circuit voltage of the fully configured battery pack, recording voltage, current, and temperature through three charge/discharge cycles, measuring the voltages at which overcharge and over-discharge protection activated and deactivated, and measuring mass before and after undergoing vacuum. Given availability of the equipment to perform these tests and measurements, it took approximately two weeks to complete this testing for each battery pack. Much of that time was dedicated to setup, waiting for charge cycles to complete, and interpretation of the results. Additional tests were required for astronaut safety on the ISS, but we would consider them to be extraneous for non-ISS CubeSat missions. These include measuring of the physical dimensions of each battery, measuring the closed circuit voltage of the fully configured battery pack, measuring the time to trigger short-circuit protection and maintaining the short for three hours to verify the protection remains enabled, and doing a dedicated vibration test at five frequencies and strengths up to 9.65 $g_{rms}$ on all three axes, with voltage measurements between each axis. These additional tests took several weeks of additional time and planning, particularly in the design, manufacturing, and modification of components to support vibration testing.

**F. Importance of a Second CubeSat Unit**

The fabrication of two identical sets of hardware in parallel is much less expensive than the same development in series, particularly if the start of the development for the second set is delayed by months or years. Small projects tend to have less stringent requirements on documentation, so details can be forgotten and lost in the time between two sets of flight hardware developed in series. Having two sets of hardware enables the development and testing of flight software while other activities proceed in parallel. Importantly, parallel development also enables the replacement of a subsystem if a problem is found, which is critical when schedules are tight. This was the case for MinXSS when the battery heater short was discovered in FM-1 at the initial pump-down for its thermal vacuum test. We were delayed half a day to swap the battery pack out with FM-2, which did not have the same issue, as compared to the weeks of delay that would have been introduced if an entirely new battery pack had to be assembled and tested. Finally, having a second flight unit allows for debugging of hardware and software after delivery and launch of the first flight unit.

**G. Low-cost Mitigation of Radiation Issues for Electronics**

The CubeSats developed at CU and LASP have generally used industrial-grade (automobile) electronic parts because those parts have wider operating temperature ranges. Typically, the automobile-grade ICs cost $10 as compared to $2 for standard commercial ICs, but this additional cost is outweighed by the significant benefits of the higher-grade components. For example, the number of uncorrupted SD card write cycles can be improved by a factor of 10–100, and the operational temperature range expanded by purchasing a $70 4 GB hardened SD-card instead of a $4 standard SD-card. The total cost impact on MinXSS for these industrial-grade electronics parts is only a few thousand dollars, a small fraction of the total budget, but it significantly improves the potential for a longer mission life. While our intention was to have electronics that could operate over a wider temperature range, automobile-grade parts may also help with radiation tolerance of the electronics. Two MinXSS prototype CDH boards were radiation tested, one to 10 kRad and another to 25 kRad; both boards survived. It is not clear if industrial-grade parts made a difference or not for passing the harder radiation test; nonetheless, it is only a small cost increment to use the higher-grade parts.

## V. Conclusions

CubeSat technologies and capabilities are now sufficiently mature to enable peer-review-journal quality science missions. This was clearly proven with the CSSWE CubeSat, which has 17 such articles to date [1, 2, 30-44] . Leveraging that success and the recent development of a commercially available, precision 3-axis ADCS, MinXSS will push the boundary of what science is possible with a CubeSat further still. The primary science objective of MinXSS is to fill a critical spectral gap in solar measurements currently made by large satellite missions at 1/100th their typical cost. All standard satellite subsystems are present in MinXSS, except propulsion, packaged in a volume that can fit in a breadbox. Many of these subsystems were custom developed by CU and LASP (e.g., CDH, EPS, SPS & XS, structure), primarily by graduate students with professional mentorship, and other subsystems were purchased from commercial vendors (e.g., flight radio, ADCS, primary science instrument).

In the future, 6U, 12U, and 27U CubeSat standards will open up even more science capabilities by allowing for larger and more sophisticated instruments. Standardized buses that can be commercially procured are now becoming available. Typically 1–2U in size, this leaves ample mass and volume to be used for the science payload. As X-band transmitters and LEO-accessible global network communications, such as GlobalStar, become available in the near-term, it will also be possible to expand data downlink capabilities. This increase in data volume is a critical need for science that involves imaging, as even a single image from a small camera would take hours to downlink at the CSSWE / MinXSS rate of 9600 bps. We note that active pixel CMOS array detectors provide one alternative mitigation strategy for this, if the entire image does not need to be downlinked. Finally, CubeSats will enable science that was not conceivable with large, monolithic spacecraft. For the same cost as a NASA Small Explorer, a constellation of dozens of CubeSats could be put into orbit to obtain simultaneous measurements over a wide spatial distribution. These novel data will enable new scientific observational analyses and provide new constraints to physical and empirical models.

## VI. Acknowledgements

This work was supported by NASA grant NNX14AN84G and NSF grant AGSW0940277, as well as the University of Colorado at Boulder Aerospace Engineering Sciences department. We would like to specially thank the many professional scientists and engineers that provided feedback at reviews and mentorship to students.